\documentclass[final]{jfm}
\usepackage{float,tabularx,graphicx}
\usepackage{amsmath,amsfonts,amsgen,amssymb,amsbsy,amstext}
\usepackage{natbib}

\title[]{On self-similarity properties of isotropic
  turbulence in numerical simulations of the compressible Euler equations}
\author[W. Schmidt, W. Hillebrandt and J. C. Niemeyer]
       {W\ls O\ls L\ls F\ls R\ls A\ls M\ns S\ls C\ls H\ls M\ls I\ls D\ls T$^{1,2}$,\ns
        W.\ns  H\ls I\ls L\ls L\ls E\ls B\ls R\ls A\ls N\ls D\ls T$^1$\\ \and 
        J.\ns C.\ns  N\ls I\ls E\ls M\ls E\ls Y\ls E\ls R$^2$\ns}
\affiliation{$^1$Max-Planck-Institut f\"ur Astrophysik, Garching, Germany\\[\affilskip]
             $^2$Universit\"at W\"urzburg, W\"urzburg, Germany}
\date{June 2004}

\newcommand{\vect}[1]{\ensuremath{\boldsymbol{#1}}}
\newcommand{\dd}{\ensuremath{\mathrm{d}}}
\newcommand{\DD}{\ensuremath{\mathrm{D}}}

\DeclareMathAlphabet{\mathsfsl}{OT1}{cmss}{m}{sl}

\begin{document}

\maketitle

\begin{abstract}

We present numerical calculations of the parameters $C_{\nu}$,
$C_{\epsilon}$ and $C_{\kappa}$ associated with the common closures
for turbulence production, dissipation and diffusion. In the case of
homogeneous and isotropic turbulence, these parameters are expected to
be statistically scale-invariant within the inertial subrange. In
order to scrutinise this conjecture, we utilised a generalisation of
the Germano filtering formalism, which is applicable to compressible
flows as well.  The filtering of data obtained from three-dimensional
direct numerical simulations of forced isotropic turbulence with Mach
numbers in the range $\sim 0.1\ldots1$ then yielded values of the
closure parameters associated with different length scales.  The
results indicate that the closure parameters are nearly universal for
subsonic or moderately transonic flows, although the resolution of
$432^{3}$ grid cells in our simulations is not quite sufficient to
clearly establish scale invariance. In addition, it was found that the
customary assumption of a kinetic Prandtl number of about unity for
the gradient-diffusion closure is flawed due to the misalignment
between turbulent flux and the gradient of the turbulence
energy. Nevertheless, sound correlation can be achieved if the flux
magnitude rather than the flux vector is locally matched. This
conclusion is particularly useful for the family of subgrid scale
models based on the turbulence energy equation.  Furthermore, the
parameter of production $C_{\nu}$ was computed in the fashion of
dynamical procedures. Thereby, superior agreement between modelled and
explicitly evaluated turbulence stresses in comparison to the
eddy-viscosity closure with constant $C_{\nu}$ was verified.

\end{abstract}

\section{Introduction}

The notion of self-similarity plays an important role in the numerical
computation of turbulent flows, particularly, in large-eddy
simulations \cite*[LES; see][]{Sagaut}.  The key idea is that the
large-scale features of the flow are shaped by the action of external
forces and the imposed boundary conditions, whereas structures on
smaller scales are mostly determined by the non-linear transfer of
kinetic energy through a cascade of vortices. The dynamics on the
largest scales, which constitute the \emph{energy-containing range},
must be explicitly computed in a LES. On the other hand, turbulent
velocity fluctuations on smaller scales, which are contained in the
so-called \emph{inertial subrange}, are considered to be statistically
self-similar.  Usually, this property of inertial-range turbulence is
expressed in terms of scale-invariant probability distribution
functions associated with structural properties of the flow. It
becomes beautifully manifest in the $k^{-5/3}$ power law of
Kolmogorov's theory. In the context of \emph{closures} for second or
higher order moments of the velocity field (and, possibly, other
quantities), the associated parameters are expected to become
asymptotically scale-invariant, once the cutoff length scale
sufficiently penetrates the inertial subrange. If this supposition was
validated, it would bear important consequences on the treatment of
the unresolved fraction of turbulence in LES, i.~e., the problem of
subgrid scale (SGS) modelling \cite*[see][]{MeneKatz00}.

Inspired by the turbulence energy equation model \cite*[see][\S
  4.3.2]{Sagaut}, we attempted to explicitly evaluate the rate of
production and dissipation, respectively, as well as turbulent
diffusion in consecutive bands of wave numbers, using data obtained
from numerical simulations of forced isotropic turbulence.  The
dissipation of kinetic energy on the smallest resolved scales in these
simulations is entirely due to the implicit viscosity produced by the
numerical scheme. The physical viscosity of the fluid is negligible.
From structural properties of the flow on larger scales, which appear
to be fairly independent of the details of dissipation, we were able
to compute the parameters $C_{\nu}$ and $C_{\kappa}$, stemming from
the eddy-viscosity and the gradient-diffusion closure,
respectively. For the rate of dissipation, we invoked the hypothesis
of local equilibrium and computed the parameter $C_{\epsilon}$
associated with the simple dimensional closure.  Furthermore, the
correlation between modelled and explicitly calculated quantities was
investigated. The numerical simulations which provided the fundamental
data are briefly outlined in \S\ref{sc:isoturb}. The underlying
formalism of filtered quantities is introduced in \S\ref{sc:flt}. The
following \S\ref{sc:prod}, \S\ref{sc:diss} and~\S\ref{sc:diff} cover
the discussion of various results for the closure parameters and
comparisons between the corresponding modelled and actual
quantities. In the conclusion, the question whether we can indeed
speak of self-similarity properties inferred form simulation data is
posed. The analysis presented in this paper points towards a positive
answer. However, more computing power in order to achieve higher
resolution would be desirable. This becomes particularly clear from
samples of the turbulence energy spectrum function discussed in
\S\ref{sc:isoturb}. In particular, the spectra display the notorious
\emph{bottleneck effect}, which is a deviation from inertial-range
scaling on length scales close to the cutoff spoils the full
visibility of self-similarity to some extent \cite*[see][]{DobHau03}.

\section{Forced Isotropic Turbulence}

\label{sc:isoturb}

In order to produce numerical realisations of isotropic turbulence, we
computed the flow of a compressible fluid set into motion by a random
force field in a cubic domain subject to periodic boundary
conditions. In the case of statistically isotropic and stationary
forcing, the outcome is an almost perfect realisation of the
mathematical paradigm of homogeneous and isotropic turbulence. Only
the imposed periodicity in Cartesian coordinates induces an inherent
anisotropy on large scales. The dynamics of the fluid is generally
determined by the following set of hydrodynamical conservation laws:
\begin{align}
  \label{eq:isoturb_cont}
  \frac{\partial}{\partial t}\rho + 
    \frac{\partial}{\partial x_{i}}\rho v_{i} & = 0,\\
  \label{eq:isoturb_momt}
  \frac{\partial}{\partial t}\rho v_{i} +
    \frac{\partial}{\partial x_{k}}\rho v_{i}v_{k} & =  
    -\frac{\partial}{\partial x_{i}}P + \rho f_{i} + 
    \frac{\partial}{\partial x_{k}}\sigma_{ik}, \\
  \label{eq:isoturb_energy}
  \frac{\partial}{\partial t}E +
    \frac{\partial}{\partial x_{k}}E v_{k} & =
    \rho f_{k}v_{k}.
\end{align}
Energy is injected into the fluid through a mechanical force field
$\vect{f}(\vect{x},t)$ called the \emph{driving force} and ultimately
dissipated by the microscopic viscosity $\nu$.  The corresponding
dissipation tensor $\sigma_{ik}$ is proportional to the local
\emph{rate of strain} of the velocity field:
\begin{equation}
  \sigma_{ik} = 2\rho\nu S_{ik}^{*} \equiv
  2\rho\nu\left(S_{ik} - \frac{1}{3}d\delta_{ik}\right),
\end{equation}
where $S_{ik}=\frac{1}{2}(v_{i,k}+v_{k,i})$ and $d=v_{i,i}$. Since
mostly velocity fluctuations on the smallest length scales contribute
to the strain, microscopic viscous dissipation becomes negligible on
larger scales and the fluid dynamics is dominated by non-linear
turbulent interactions. At sufficiently high resolution, the cutoff
arising from the discretisation in numerical simulations falls into
this very range of scales. In consequence, either a subgrid-scale
model has to be employed in order to account for the energy transfer
from numerically resolved towards unresolved length scales, or a
dissipative finite-volume scheme is applied, which properly smoothes
the flow on the smallest resolved scales.  We chose the latter
approach and adopted the \emph{piece-wise parabolic method} (PPM)
proposed by \cite*{CoWood84} for the solution of the hydrodynamical
equations. The idea of employing the PPM for direct numerical
simulations of turbulent flows was particularly advocated by
\cite*{SyPort00}, following systematic convergence tests and
comparisons to conventional methods with explicit treatment of the
viscosity term. From the numerical point of view, the treatment of the
hydrodynamical equations with the PPM corresponds to the limit of zero
physical viscosity. In essence, the compressible Euler equations are
solved, with the velocity fluctuations on scales smaller than the grid
resolution $\Delta$ being damped out by numerical dissipation.

The driving force $\vect{f}(\vect{x},t)$ is composed in spectral
space, using a three-dimensional generalisation of the scalar
\emph{Ornstein-Uhlenbeck process}, as proposed by \cite{EswaPope88}.
The evolution of the Fourier transform $\hat{\vect{f}}(\vect{k},t)$ is
given by the following Langevin-type stochastic differential equation:
\begin{equation}
  \label{eq:stirr_mode_evol}
  \dd\hat{\vect{f}}(\vect{k},t) = 
  -\hat{\vect{f}}(\vect{k},t)\frac{\dd t}{T} +
  F_{0}\sum_{jlm}\left(\frac{2\sigma^{2}(\vect{k})}{T}\right)^{1/2}
    \delta(\vect{k}-\vect{k}_{jlm})
    \boldsymbol{\mathsfsl{P}}_{\zeta}(\vect{k})\cdot\dd\vect{\mathcal{W}}_{t},
\end{equation}
The second term on the right hand side accounts for a random diffusion
process, which is constructed from a three-component \emph{Wiener
process} $\vect{\mathcal{W}}_{t}$. The distribution of each component
is normal with zero mean and variance $\dd t$. The wave vectors
$\vect{k}_{jlm}$ are dual to the position vectors of the cells in the
numerical discretisation of the fundamental domain.  The symmetric
tensor $\boldsymbol{\mathsfsl{P}}_{\zeta}(\vect{k})$ is defined by the linear
combination of the projection operators perpendicular and parallel to
the wave vector. The components of
$\boldsymbol{\mathsfsl{P}}_{\zeta}(\vect{k})$ can be expressed as
\begin{equation}
  (P_{ij})_{\zeta}(\vect{k}) = 
  \zeta P_{ij}^{\perp}(\vect{k}) + (1-\zeta)P_{ij}^{\parallel}(\vect{k}) =
  \zeta\delta_{ij} + (1-2\zeta)\frac{k_{i}k_{j}}{k^{2}},
\end{equation}
where the spectral weight $\zeta$ determines whether the resulting
force field in physical space is purely solenoidal, dilatational or a
combination of both. The variance $\sigma^{2}(\vect{k})$ specifies the
spectrum of the force field. We use a quadratic function, which
confines the modes of the force to a narrow interval of wavenumbers,
$k\in[0,2k_{0}]$. The wave number $k_{0}$ determines the
\emph{integral length scale} of the flow, $L=2\pi/k_{0}$.

The root mean square of the specific driving force is determined by
the characteristic magnitude $F_{0}$ and the weight $\zeta$:
\begin{equation}
  \label{eq:stirr_rms}
  f_{\mathrm{rms}} = 
  \sum_{jlm}\langle\hat{\vect{f}}_{jlm}(t)\cdot\hat{\vect{f}}_{jlm}(t)\rangle
  \simeq (1 - 2\zeta + 3\zeta^{2})F_{0}^{2}.
\end{equation}
Since $F_{0}$ has the physical dimension of acceleration, it can be
expressed as the \emph{characteristic velocity} $V$ of the flow
divided by the \emph{integral time scale}, which is given by the
auto-correlation time $T$ of the driving force
(\ref{eq:stirr_mode_evol}). Setting $T=L/V$, we have $F_{0}=V/T=L
V^{2}$, and, starting with a homogeneous fluid at rest, the flow is
developing towards a fully turbulent steady state within about two
integral time scales.

\begin{table}
  \begin{center}
    \begin{tabular}{@{}llrrr@{}}
      $\zeta$ & $V/c_{0}$ & $t_{\mathrm{d}}/T$ & $t_{\mathrm{f}}$/T & $N_{\Delta t}$ \\[3pt]
      1.0  & 0.084 &     & 2.5  & 5815 \\
      1.0  & 0.42  & 3.0 & 8.0  & 6343 \\
      0.75 & 0.66  & 5.0 & 10.0 & 6351 \\
      0.2  & 1.39  & 5.0 & 10.0 & 3356 \\
    \end{tabular}
  \end{center}
  \caption{List of simulation parameters: spectral weight $\zeta$ of
  the driving force, characteristic Mach number $V/c_{0}$, onset of
  decay $t_{\mathrm{d}}/T$, end of simulation $t_{\mathrm{f}}/T$,
  number of time steps $N_{\Delta t}$. }
  \label{tb:dns}
\end{table}

As our work was motivated by the problem of turbulent burning
processes in thermonuclear supernovae, we applied the equation of
state for a degenerate electron gas in combination with non-degenerate
nuclei. This form of matter occurs in compact stellar remnants called
white dwarfs \cite*[see][]{HilleNie00, Rein01}. However, once
quantities are scaled in terms of characteristic parameters, no major
differences to turbulence in, say, an ideal gas are found. The
parameters chosen for several particular simulations are listed in
table~\ref{tb:dns}. In the simulations with $\zeta=1$, turbulence was
produced by purely solenoidal forcing or ``stirring''. We chose two
different values of the characteristic Mach number $V/c_{0}$, where
$c_{0}$ is the speed of sound in the initially homogeneous fluid at
rest. For the lower value, the flow is completely subsonic, whereas
for the higher value, it is marginally transonic. Furthermore, we run
two simulations with partially dilatational forcing, i.~e., the force
field has compressive components. In both cases, the velocity of the
flow locally exceeds the speed of sound and shocklets are formed. For
all simulations, $432^{3}$ grid cells were used and $\alpha=X/L=3$,
where $X$ is the linear size of the fundamental domain. The evolution
was computed over an elapse of several integral time scales, such that
data dumps corresponding to a statistically stationary state with
energy injection balanced by dissipation were obtained. Finally, decay
phases were initiated at a certain time $t_{\mathrm{d}}$ by inhibiting
the random increments in the evolution of the stochastic force field.
Also included in table~\ref{tb:dns} are the total durations
$t_{\mathrm{f}}$ of the simulations in units of the corresponding
integral time scales and the required total number of time steps.

\begin{figure}
  \begin{center}
    \includegraphics[width=\linewidth]{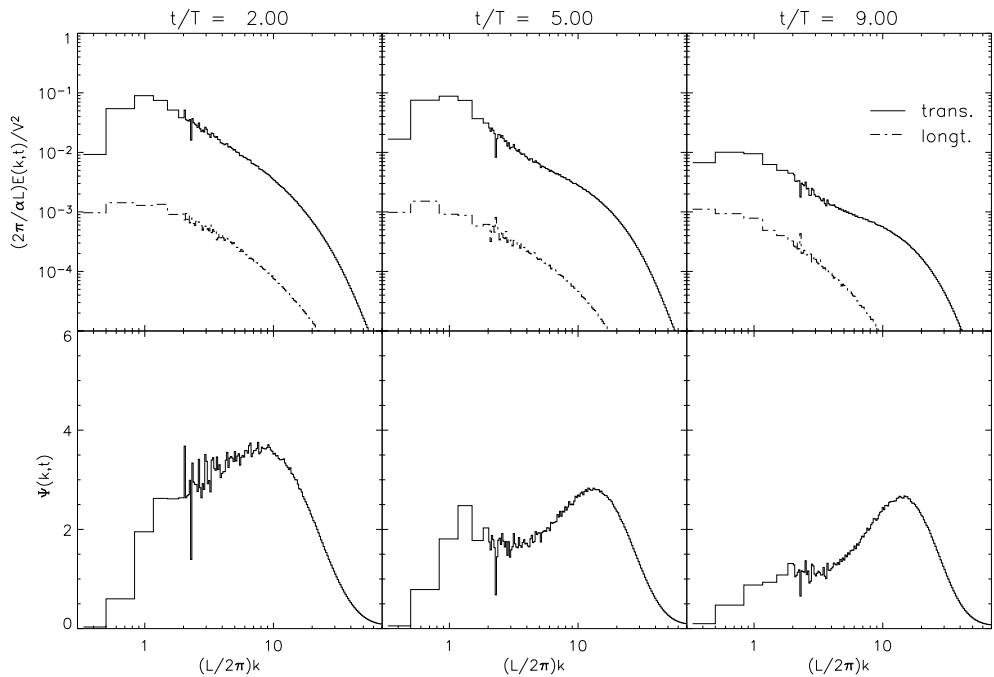}
    \caption{ Plots of the transversal and longitudinal parts of the
      turbulence energy spectrum function $\tilde{E}(k,t)$ (a--c: top
      panels) as well as the compensated transversal spectrum function
      $\Psi^{\perp}(k,t)$ (d--f: bottom panels) at different times in
      a simulation of forced isotropic turbulence. The characteristic
      Mach number is $V/c_{0}=0.66$ and the spectral weight of the
      driving force $\zeta=0.75$. }
    \label{fg:spect075}
  \end{center}
\end{figure}

Samples of the normalised energy spectrum function
$\tilde{E}(\alpha\tilde{k},t)=(2\pi/\alpha LV^{2})E(k,t)$ at
representative stages in the production regime, in statistical
equilibrium and in the advanced decay regime are plotted in
figure~\ref{fg:spect075} (a--c) for the simulation with
$\zeta=0.75$. Due to the discrete nature of the numerical data, the
kinetic energy of modes within certain wave number bins was summed
up. Also shown are the longitudinal and a transversal fractions of the
energy spectrum functions corresponding to, respectively, the
compressible and incompressible components of the velocity field. 
Figure~\ref{fg:spect075} (d--f) shows the corresponding plots of the
\emph{compensated} solenoidal spectrum function
\begin{equation}
  \label{eq:comps_spect}
  \Psi^{\perp}(t) =
  \left[\frac{\alpha}{2\pi}\langle\tilde{\epsilon}(t)\rangle\right]^{-2/3}
    (\alpha\tilde{k})^{5/3}\tilde{E}^{\perp}(t),  
\end{equation}
where $\langle\tilde{\epsilon}\rangle=(T/V^{2})\langle\epsilon\rangle$
is the normalised mean rate of dissipation. In the inertial subrange,
$\Psi_{\mathrm{s}}(t)\simeq C$ is expected. $C$ is known as the
Kolmogorov constant and considered to be more or less universal.  From
the spectra shown in figure~\ref{fg:spect075}, it becomes clear that a
resolution of $N=432$ is just at the brink where an inertial subrange
begins to take shape.  Even in the quasi-equilibrium state, there is
merely a narrow window of wave numbers in the vicinity of
$\tilde{k}=(L/2\pi)k=3.0$, in which nearly Kolmogorov scaling with
$C\approx 1.7$ is found. In fact, this value of the Kolmogorov
constant is comfortably within the range of numerical results reported
in the literature \cite*[see][]{YeuZhou97}.  Both in
figure~\ref{fg:spect075}~(e) and~\ref{fg:spect075}~(f), there is a
pronounced maximum of the compensated spectrum function at
$\tilde{k}\approx 15$ corresponding to a flattening of the energy
spectrum in comparison to the Kolmogorov law
(figures~\ref{fg:spect075}~b and~\ref{fg:spect075}~c). This so-called
\emph{bottleneck effect} was observed in many numerical simulations
\cite[see][]{DobHau03}. At time $\tilde{t}=t/T=1.5$, on the other
hand, small-scale features have only partially developed and the
spectrum does not show a bump at higher wave numbers
(figures~\ref{fg:spect075}~a and~\ref{fg:spect075}~d). We will not
further discuss the bottleneck effect here, but take it as a genuine
feature of our simulations.

\section{Hierarchical filtering}

\label{sc:flt}

In this section we shall elaborate the formal concepts of filters and
physical fields smoothed over the associated length scales.  The
formalism was first introduced by \cite{Germano92} in the framework of
incompressible flows. In the following, \emph{mass-weighted} filtered
quantities are defined, which are applicable to compressible flows as
well. Remarkably, one of the key results obtained by Germano, the
so-called \emph{Germano identity}, carries over to the compressible
case without modification.  In order to avoid confusion, we will
subsequently write the quantities given by the hypothetical exact
solution of the dynamical equations in the continuum limit, i.\ e.,
for infinite resolution, with the superscript $\infty$ set on top of
the corresponding symbol, whereas the standard symbols refer to
numerically computable quantities.  Notwithstanding the deplorable
lack of any proof of existence, the former shall be called the
\emph{ideal} quantities\footnote{ According to Plato's Cave Analogy,
the objects we experience in the world are related to their
\emph{ideals} like the silhouettes of outside things to someone
trapped in a dark cave.  In the same way, one can think of a
numerically simulated flow as being merely a vague image of the
corresponding physical flow or exact mathematical solution of the
equations of motion.  }. For instance,
$\overset{\infty}{\vect{v}}(\vect{x},t)$ is identified with the ideal velocity
field, i.~e., the exact solution of the Navier-Stokes or Euler
equation, while it is understood that $\vect{v}(\vect{x},t)$ is an
approximation to the velocity field, say, an interpolating function to
some finite-volume solution.

Since we are mostly concerned with statistically stationary
homogeneous turbulence, it is convenient to introduce an infinite
series of homogeneous and time-independent filters $\langle\
\rangle_{n}$. The kernel of the $n$-th filter is denoted as
$G_{n}(\vect{x})$. The corresponding \emph{filter operation} on a
dynamical quantity $q(\vect{x},t)$ is defined by
\begin{equation}
  \label{eq:q_flt_n}
  q^{(n)}(\vect{x},t) = \int\dd^{3}x'\,G_{n}(\vect{x}')\overset{\infty}{q}(\vect{x}',t)
\end{equation}
and symbolically written as $q^{(n)}=\langle\overset{\infty}{q}\rangle_{n}$.  
In the limit $n\rightarrow\infty$, the identity operator is obtained.
Here we shall assume that $\langle\ \rangle_{n}$ is a Gaussian filter
with a characteristic length $\Delta_{n}$ and wave number
$k_{n}=\pi/\Delta_{n}$.  The corresponding kernel is given by 
\cite*[see][ \S13.2]{Pope}
\begin{equation}
  G_{n}(\vect{x}) = \left(\frac{6}{\pi\Delta_{n}^{2}}\right)^{3/2}
    \exp\left(-\frac{6|\vect{x}|^{2}}{\Delta_{n}^{2}}\right).
\end{equation}
In the following, a series of Gaussian filters with $\Delta_{n}=L/n$,
$n\in\mathbb{N}$, is used, where $L$ is the integral length scale of the
flow.

The velocity field $\vect{v}^{[n]}(\vect{x},t)$ at the $n$-th filter
level is defined by a \emph{Favre mass-weighted} filter operation:
\begin{equation}
  \label{eq:vel_flt_n}
  \vect{v}^{[n]}(\vect{x},t) = \frac{1}{\rho^{(n)}(\vect{x},t)}
  \int\dd^{3}x'\,G_{n}(\vect{x}')\overset{\infty}{\rho}(\vect{x}',t)
                                 \overset{\infty}{\vect{v}}(\vect{x}',t)
\end{equation}
or, for brevity, 
$\vect{v}^{[n]}=\langle\overset{\infty}{\rho}\overset{\infty}{\vect{v}}\rangle_{n}/
\langle\overset{\infty}{\rho}\rangle_{n}$.
Filtering twice, we set
\begin{equation}
  \vect{v}^{[m][n]}\rho^{(m)(n)} = 
  \langle\rho^{(m)}\vect{v}^{[m]}\rangle_{n} =
  \langle\langle\overset{\infty}{\rho}\overset{\infty}{\vect{v}}\rangle_{m}\rangle_{n},
\end{equation}
where $\rho^{(m)(n)}=\langle\langle\overset{\infty}{\rho}\rangle_{m}\rangle_{n}$.
If two filters of characteristic length scales $\Delta_{m}$ and
$\Delta_{n}\gg\Delta_{m}$ are operating in succession, we have
\begin{equation}
  \vect{v}^{[m][n]}\rho^{(m)(n)} =
  \langle\rho^{(m)}\vect{v}^{[m]}\rangle_{n} \simeq 
  \langle\overset{\infty}{\rho}\overset{\infty}{\vect{v}}\rangle_{n} =
  \vect{v}^{[n]}\rho^{(n)}.
\end{equation}
The validity of this approximation becomes immediately clear
from the product of the kernels of the Gaussian filters in spectral
space \cite*[see][\S13.2]{Pope}:
\begin{equation}
  \hat{G}_{m}(k)\hat{G}_{n}(k) = 
  \exp\left[-\frac{k^{2}(\Delta_{m}^{2}+\Delta_{n}^{2})}{24}\right] \simeq 
  \exp\left[-\frac{k^{2}\Delta_{n}^{2}}{24}\right] = \hat{G}_{n}(k).
\end{equation}
Because the convolution with the filter kernel in physical space
corresponds to the multiplication of the Fourier transforms in
spectral space, it follows that 
$\langle\langle q\rangle_{m}\rangle_{n}\simeq\langle q\rangle_{n}$ if
$\Delta_{n}\gg\Delta_{m}$.

The definition of the filtered velocity at the level $n$
 (\ref{eq:vel_flt_n}) is particularly useful, because the equation of
 motion for the fluid on scales $l\gtrsim\Delta_{n}$ is given by a
 \emph{quasi-Navier-Stokes equation}:
\begin{equation}
  \label{eq:qnse}
  \frac{\partial}{\partial t}\,\rho^{(n)} v_{i}^{[n]} +
  \frac{\partial}{\partial x_{k}}\,\rho ^{(n)}v_{i}^{[n]}v_{k}^{[n]}=
  -\frac{\partial P^{(n)}}{\partial x_{i}} + \rho^{(n)}f_{i} + 
  \frac{\partial}{\partial x_{k}}(\sigma_{ik}^{(n)}+ \tau_{ik}^{[n]}).
\end{equation}
The specific force $f_{i}$ stirring the fluid is assumed to vary on
the largest scales only.  This is why the force field is only marginally
affected by the filtering operation, and one can set $f_{i}^{[n]}\simeq
f_{i}$ for $n\gg 1$. In comparison to the \emph{physical} Navier-Stokes equation,
\begin{equation}
  \label{eq:nse}
  \frac{\partial}{\partial t}\,
    \overset{\infty}{\rho}\overset{\infty}{v_{i}} +
  \frac{\partial}{\partial x_{k}}\,
    \overset{\infty}{\rho}\overset{\infty}{v_{i}}\overset{\infty}{v_{k}}=
  -\frac{\partial \overset{\infty}{P}}{\partial x_{i}} + \rho f_{i} + 
  \frac{\partial}{\partial x_{k}}\sigma_{ik},
\end{equation}
the viscous stress tensor $\overset{\infty}{\sigma_{ik}}=
2\overset{\infty}{\rho}\nu\overset{\infty}{S_{ik}^{\ast}}$ for a fluid
of microscopic viscosity $\nu$ is enhanced by the \emph{turbulence
stress tensor} associated with the $n$-th filter\footnote{Here the
opposite sign as customary in most of the literature is used in order
to make $\tau_{ik}$ a proper stress tensor, which enters the
right-hand side of the equation for the resolved energy with positive
sign.}.  The latter is defined by
\begin{equation}
    \label{eq:turb_stress}
    \tau^{[n]}_{ik} =
    -\langle\overset{\infty}{\rho}\overset{\infty}{v_{i}}
                                  \overset{\infty}{v_{k}}\rangle_{n}
    + \rho^{(n)}v_{i}^{[n]}v_{k}^{[n]}.
\end{equation}
In particular, if $\Delta_{n}$ is large compared to the length scale
of viscous dissipation, then $\tau^{[n]}_{ik}$ dominates over
$\sigma_{ik}^{(n)}$, and (~\ref{eq:qnse}) becomes a \emph{quasi-Euler
equation}, in which the microscopic viscosity $\nu$ does not appear at
all.

For two consecutive filter levels, say, $n$ and $n-1$, there is
a simple algebraic relation between the corresponding stress tensors:
\begin{equation}
  \label{eq:germano}
  \tau^{[n][n-1]}_{ik} = 
  \langle\tau^{[n]}_{ik}\rangle_{n-1} +
  \tau^{[n,n-1]}(v_{i}^{[n]},v_{k}^{[n]}),
\end{equation}
where
\begin{equation}
  \begin{split}
  \tau^{[n][n-1]}_{ik} =
  &-\langle\langle\overset{\infty}{\rho}\overset{\infty}{v_{i}}
                                \overset{\infty}{v_{k}}\rangle_{n}\rangle_{n-1}
  + \frac{\langle\langle\overset{\infty}{\rho}\overset{\infty}{v_{i}}\rangle_{n}\rangle_{n-1}
          \langle\langle\overset{\infty}{\rho}\overset{\infty}{v_{k}}\rangle_{n}\rangle_{n-1}}
         {\langle\langle\overset{\infty}{\rho}\rangle_{n}\rangle_{n-1}} = \\
  &-\langle\langle\overset{\infty}{\rho}\overset{\infty}{v_{i}}
                                \overset{\infty}{v_{k}}\rangle_{n}\rangle_{n-1}
  + \rho^{(n)(n-1)}v_{i}^{[n][n-1]}v_{k}^{[n][n-1]}.
  \end{split}
\end{equation}
and
\begin{equation}
  \tau^{[n,n-1]}(v_{i}^{[n]},v_{k}^{[n]}) =
  -\langle\rho^{(n)}v_{i}^{[n]}v_{k}^{[n]}\rangle_{n-1} + 
  \frac{1}{\langle\rho^{(n)}\rangle_{n-1}}
    \langle\rho^{(n)}v_{i}^{[n]}\rangle_{n-1}
    \langle\rho^{(n)}v_{k}^{[n]}\rangle_{n-1}
\end{equation}
is the \emph{intermediate} stress tensor at the filter level $n$
associated with the velocity field filtered at the level $n-1$.  The
above relation was originally formulated by \cite{Germano92} for
incompressible flows. The relation also applies to arbitrary filter
levels, $m$ and $n$, say. In the limit $\Delta_{n}\gg\Delta_{m}$, the
contribution from $\langle\tau^{[m]}_{ik}\rangle_{n}$ becomes
negligible and
\begin{equation}
  \label{eq:turb_stress_asympt}
  \tau^{[n]}_{ik} \simeq
  \tau^{[m][n]}_{ik} \simeq \tau^{[m,n]}(v_{i}^{[m]},v_{k}^{[m]}),
\end{equation}
i.\ e., the turbulence stress associated with the scale $\Delta_{n}$
is not sensitive to the flow structure on much smaller scales.  The
asymptotic limit of the Germano relation is especially useful for the
numerical evaluation of turbulence stresses from numerical data.
 
Utilising the \emph{consistent Germano decomposition}
\cite*[see][\S3.3.2]{Sagaut}, the turbulence energy associated with
length scales $l\lesssim\Delta_{n}$ is given by the difference between
the filtered and the resolved kinetic energy at the filter level
$n$. This particular notion of turbulence energy was introduced by
\cite{Germano92} and is called the \emph{generalised turbulence
energy}.  At any filter level, the energy is readily defined by the
trace of turbulence stress tensor:
\begin{equation}
  K^{[n]}=\rho^{(n)}k^{[n]}=-\frac{1}{2}\tau^{[n]}_{ii}. 
\end{equation}
This definition avoids several difficulties stemming from the
mathematical properties of common filter operations, if the turbulence
energy is defined in terms of velocity fluctuations relative to the
filtered velocity field in the alternative Leonard's decomposition
\cite*[see][\S3.3.1]{Sagaut}.  Contracting the compressible Germano
identity~(\ref{eq:germano}), we obtain the relation
\begin{equation}
  K^{[n][n-1]} = \langle K^{[n]}\rangle_{n-1} + K^{[n,n-1]},
\end{equation}
where
\begin{equation}
  K^{[n,n-1]} = -\frac{1}{2}\tau^{[n,n-1]}(v_{i}^{[n]},v_{i}^{[n]})
\end{equation}
is the intermediate turbulence energy, i.\ e., the kinetic energy of
modes which are concentrated in the spectral band
$[\pi/\Delta_{n-1},\pi/\Delta_{n}]$. As opposed to the spectral
filter, there are no sharp boundaries between adjacent bands
associated with Gaussian filters. Nevertheless, the notion of
turbulence energy as proposed by Germano is well-defined and
unambiguously associates some energy contents with each band of
wave numbers. 

The scale separation of the energy budget by means of filtering on a
characteristic scale $\Delta_{n}$ also yields a conservation law for
the turbulence energy $K^{[n]}$. The result is completely analogous to
the dynamical equation for SGS turbulence energy and it also entails
various closure problems \cite[see][\S3.3.2]{Germano92,Sagaut}.
Adopting the standard closures for production, diffusion and
dissipation \cite[see][\S4.3.2]{Sagaut}, we obtain the following
approximate equation for the time evolution of $K^{[n]}$:
\begin{equation}
  \begin{split}
  \label{eq:turb_energy}
  \frac{\DD^{[n]}}{\DD t}k^{[n]} -& \frac{1}{\rho^{(n)}}
  \vect{\nabla}\cdot\left(\rho^{(n)}C_{\kappa}^{(n)}\Delta_{n}
    \sqrt{k^{[n]}}\vect{\nabla}k^{[n]}\right) = \\  
  & C_{\nu}^{(n)}\Delta_{n}\sqrt{k^{[n]}}|S^{\ast\,[n]}|^{2} -
  \frac{2}{3}k^{[n]}d^{[n]} -
  C_{\epsilon}^{(n)}\frac{(k^{[n]})^{3/2}}{\Delta_{n}},
  \end{split}
\end{equation}
where $\frac{\DD^{[n]}}{\DD t}=\frac{\partial}{\partial
t}+\vect{v}^{[n]}\cdot\vect{\nabla}$ is the Lagrangian time
derivative with respect to the filtered velocity field,
$S_{ik}^{\ast\,[n]}=\frac{1}{2}(\partial_{k}v_{i}^{[n]}+\partial_{i}v_{k}^{[n]})-\frac{1}{3}d^{[n]}\delta_{ik}$
are the components of the trace-free rate-of-strain tensor and
$d^{[n]}=\partial_{i}v_{i}^{[n]}$ is the divergence of the velocity field
filtered at the level $n$. Now the question of self-similarity boils
down to the scaling-behaviour of the parameters $C_{\nu}^{(n)}$,
$C_{\epsilon}^{(n)}$ and $C_{\kappa}^{(n)}$ associated with the
different filters in the hierarchy.

\section{Kinetic energy transfer}

\label{sc:prod}

The rate of transfer of kinetic energy from velocity fluctuations on
length scales $l>\Delta_{n}$ towards those on smaller scales is given by the
contraction of the turbulence stress tensor and the rate of strain
tensor at the level of the $n$-th filter:
\begin{equation}
  \Pi^{(n)} = \tau_{ik}^{[n]}S_{ik}^{[n]}.
\end{equation}
Note that $S_{ik}^{[n]}$ is the symmetrised derivate of the filtered
velocity field, i.~e.,
$S_{ik}^{[n]}=\frac{1}{2}(\partial_{k}v_{i}^{[n]}+\partial_{i}v_{k}^{[n]})$. Since
the filter operation involves mass-weighing, differentiation and
filtering do \emph{not} commute. The most common closure for the rate
of energy transfer is the \emph{eddy-viscosity closure} for the
trace-free part of the turbulence stress tensor:
\begin{equation}
  \tau_{ik}^{\ast\,[n]} \circeq 2\rho^{(n)}\nu^{(n)}S_{ik}^{\ast\,[n]},
\end{equation}
where
$\tau_{ik}^{\ast\,[n]}=\tau_{ik}^{[n]}-\frac{1}{3}\tau_{ll}^{[n]}\delta_{ik}=
\tau_{ik}^{[n]}+\frac{2}{3}K^{[n]}\delta_{ik}$, and $\nu^{(n)}$ is the
\emph{turbulent viscosity} of the fluid at the length scale
$\Delta_{n}$. Since viscosity can be expressed as the product of a
length scale and a characteristic velocity, a customary hypothesis
identifies $\nu^{(n)} = C_{\nu}^{(n)}\Delta_{n}\sqrt{k^{[n]}}$.
Hence,
\begin{equation}
  \label{eq:turb_prod}
  \Pi^{(n)} - \frac{2}{3}k^{[n]}d^{[n]} \circeq 
    \rho^{(n)}C_{\nu}^{(n)}\Delta_{n}\sqrt{k^{[n]}}|S^{\ast\,[n]}|^{2}.
\end{equation}

If turbulence were self-similar within a certain range of length
scales, then one would expect $C_{\nu}^{(n)}$ to be scale invariant,
when it was averaged over appropriate regions of the flow.  In order
to test this proposition \emph{a priori}, we filtered simulation data
in the intermediate range of scales between the grid resolution and
the integral scale. For the explicit evaluation of the turbulence
stress tensor, an \emph{enhanced viscosity approximation} was
applied. This means that the ideal velocity field in
definition~(\ref{eq:turb_stress}) is replaced by the numerically
computed field, which is smooth on scales $l\lesssim\Delta$ due to
numerical discretisation. Equivalently, one can think of the
smoothness being caused by an associated numerical viscosity, which
enhances the physical viscosity of the fluid. By the same line of
reasoning as in the case of two filters with $\Delta_{n}\gg
\Delta_{m}$, it follows that $\langle\overset{\infty}{q}\rangle_{n}
\simeq
\langle\langle\overset{\infty}{q}\rangle_{\mathrm{eff}}\rangle_{n}=
\langle q\rangle_{n}$ if $\Delta_{n}\gg \Delta_{\mathrm{eff}}$.
Moreover, $\exists N:\Delta_{N+1}<\Delta_{\mathrm{eff}}\le\Delta_{N}$.
According to the asymptotic equation~(\ref{eq:turb_stress_asympt}),
$\tau_{ik}^{[n]}$ can be approximated by $\tau_{ik}^{[N,n]}$.  This,
in turn, implies
\begin{equation}
  \label{eq:turb_stress_visc}
  \tau^{[n]}_{ik} \simeq
  -\langle\rho v_{i}v_{k}\rangle_{n} + \rho^{(n)}v_{i}^{[n]}v_{k}^{[n]}
\end{equation}
for a filter of significantly larger characteristic length than the the
numerical scale $\Delta_{\mathrm{eff}}$. Here $v_{i}$ is the PPM
solution of the quasi-Euler equation~(\ref{eq:isoturb_momt}). 

\begin{figure}
  \begin{center}
    \includegraphics[width=0.9\linewidth]{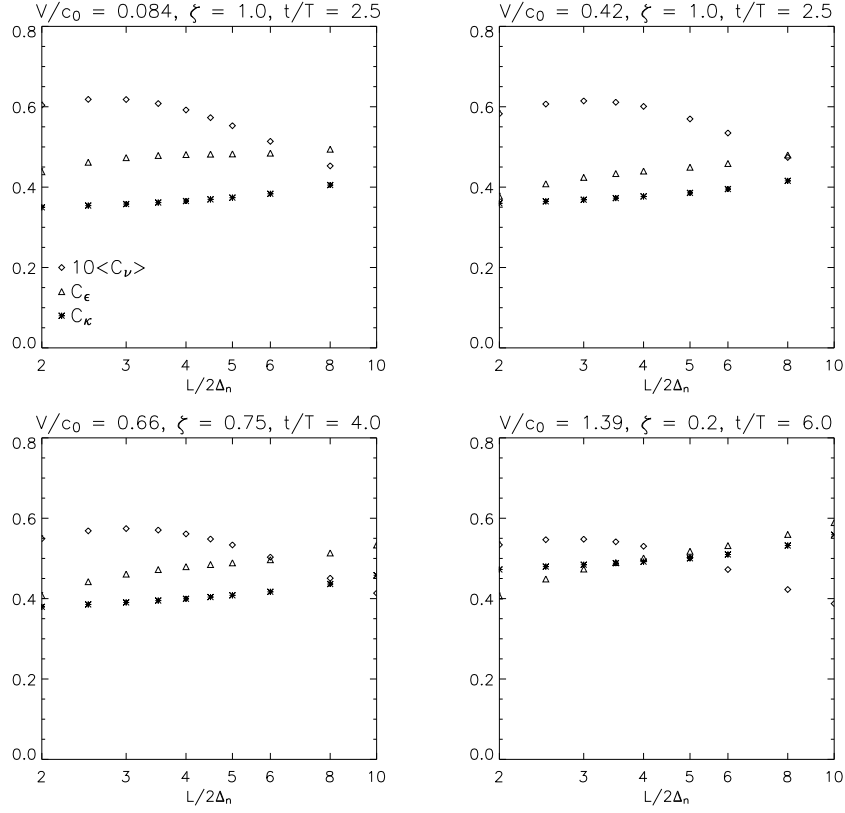}
    \caption{ Numerically evaluated closure parameters $\langle
      C_{\nu}\rangle$, $C_{\epsilon}$ and $C_{\kappa}$ as
      functions of the normalised wave number
      $\tilde{k}_{n}=L/2\Delta_{n}$ for Gaussian filters of
      characteristic length $\Delta_{n}$.  }
    \label{fg:sgs_stat}
  \end{center}
\end{figure}

However, the actual range of inertial scales in the case of the
simulation discussed in \S\ref{sc:isoturb} is rather marginal. From
the turbulence energy spectrum shown in figure~\ref{fg:spect075}, one
can see that approximate Kolmogorov scaling is found for dimensionless
wave numbers in the narrow range $2\lessapprox\tilde{k}\lessapprox
5$. The dimensionless wave number associated with a Gaussian filter of
characteristic length scale $\Delta_{n}$ is
$\tilde{k}_{n}=(\pi/\Delta_{n})(L/2\pi)= L/2\Delta_{n}$. Thus, only
filters with $0.1\lessapprox\Delta_{n}/L\lessapprox 0.25$ are more or
less suitable for calculating $C_{\nu}^{(n)}$
from~(\ref{eq:turb_prod}), if the the turbulence stress tensor is
substituted by the enhanced viscosity
approximation~(\ref{eq:turb_stress_visc}).  Notwithstanding these
tight constraints, mean values of $C_{\nu}^{(n)}$ were calculated from
a sample of data sets using several different filters.  The results
are plotted as functions of the characteristic filter wave number in
figure~\ref{fg:sgs_stat}. Although, as a consequence of the rather
limited resolution of the simulations, pronounced self-similarity is
not apparent, basic trends can be discerned.  In the case of purely
subsonic turbulence with the characteristic Mach number
$V/c_{0}\approx 0.084$, the plotted values seem to indicate a maximum
of $\langle C_{\nu}^{(n)}\rangle$ near the wave number
$\tilde{k}=2.0$, which marks the upper bound of the energy-containing
subrange. Towards higher wave numbers, i.\ e., for filters of smaller
characteristic length, $\langle C_{\nu}^{(n)}\rangle$ decreases and
eventually flattens in the vicinity of maximum dissipation at the wave
number $\tilde{k}_{\mathrm{p}}\approx 13.5$. For the other
simulations, a similar behaviour emerges, but there is seemingly a
trend toward smaller production parameters for increasing Mach number
and partially dilatational forcing.  In any case, we suggest to adopt
$\langle C_{\nu}\rangle\approx 0.06$ as a good value for fully
developed turbulence. This choice is further supported by the
selection of numerical values listed in table~\ref{tb:sgs_compr},
which were obtained with the particular filter of characteristic wave
number $\tilde{k}_{6} = 3$. Near this wave number, the compensated
energy spectra exhibit local minima.

\begin{table}
  \begin{center}
    \begin{tabular}{@{}lllccc@{}}
    $V/c_{0}$ & $\zeta$ & $t/T$ &
    $\langle C_{\nu}^{(6)}\rangle$ & $C_{\epsilon}^{(6)}$ & 
    $C_{\kappa}^{(6)}$  \\
    0.084 & 1.0 & 2.5 & 0.0618 & 0.473 & 0.358 \\
    0.42  & 1.0 & 2.5 & 0.0615 & 0.424 & 0.369 \\
    0.66 & 0.75 & 4.0 & 0.0574 & 0.461 & 0.390 \\
    1.39 & 0.20 & 6.0 & 0.0548 & 0.474 & 0.484 \\
  \end{tabular}
  \caption{Mean production, dissipation and diffusion parameters
    calculated with a Gaussian filter of characteristic length
    $\Delta_{6} = L/6 = 24\Delta$. The values in the first three
    columns specify the characteristic Mach number $V/c_{0}$, the
    spectral weight $\zeta$ of the driving force and the instant of
    time $t/T$ for the chosen data sets extracted from different
    numerical simulations. }
  \label{tb:sgs_compr}
  \end{center}
\end{table}

A different notion of self-similarity originates from the correlation
of turbulence stresses modelled upon the eddy-viscosity closure in
consecutive bands of wave numbers.  Actually, so-called dynamical
procedures for the computation of subgrid-scale closure parameters, in
particular, $C_{\nu}$, make use of this kind of
similarity. Originally, dynamical formulations of the well-known
Smagorinsky model and the SGS turbulence energy equation model
utilised the proposition that turbulence stresses at different filter
levels are similar
\cite*[see][]{GerPio91,ErleHuss92,PioLiu95,MeneKatz00}.  The ratio of
the filter length scales in this approach is usually set equal to a
factor of two.  Thus, using the notation introduced in
\S\ref{sc:flt}, the similarity hypothesis amounts to $C^{(n)} =
\langle C^{(2n)}\rangle_{n}$. Contrary to the statistical scale
invariance of $C^{(n)}$ investigated earlier in this section, this is
a \emph{local} relation, which presumes significant correlation
between the turbulence stresses $\tau_{ik}^{(n)}$ and
$\langle\tau_{ik}^{(2n)}\rangle_{n}$.

On the other hand, \cite*{LiuMen94} found evidence for a more
pronounced correlation between $\tau_{ik}^{(2n)}$ and the intermediate
stress tensor $\tau_{ik}^{[2n,n]}$. The latter is also known as the
Leonard's stress tensor in the context of SGS models. They processed
data obtained from velocity measurements in round jet experiments with
filters of varying characteristic length scale in order to evaluate
the turbulence stresses and the rate of energy transfer across certain
wave numbers. Making use of the data produced in our numerical
simulations, we re-investigated the relation reported by
\cite{LiuMen94}.  To this end, the function
$C_{\nu}^{(2n,n)}(\vect{x},t)$ was determined from the intermediated
stress tensor $\tau_{ik}^{[2n,n]}$ associated with the range of scales
$\Delta_{2n}\lessapprox l\lessapprox\Delta_{n}=2\Delta_{2n}$:
\begin{equation}
  \label{eq:c_prod_locl}
  C_{\nu}^{(2n,n)} = 
  \frac{\tau^{\ast\,[2n,n]}(v_{i}^{[2n]},v_{k}^{[2n]})S_{ik}^{[n]}}
       {\rho^{(n)}\Delta_{n}\sqrt{k^{[2n,n]}}|S^{\ast\,[n]}|^{2}}
\end{equation}
The quantities in the above expression can be evaluated by filtering
the data at two levels $n$ and $2n$, in between the energy containing
and the dissipation range.  The \emph{similarity closure} for the
turbulence stress $\tau_{ik}^{[2n]}$ associated with the length scale
$\Delta_{2n}$ is then given by
\begin{equation}
  \label{eq:prod_strong}
  \tau_{ik}^{\ast\,[2n]}S_{ik}^{[2n]} \circeq 
    C_{\nu}^{(2n,n)}\rho^{(2n)}\Delta_{2n}\sqrt{k^{[2n]}}|S^{\ast\,[2n]}|^{2}.
\end{equation}
In order to validate the modelled rate of production according
to~(\ref{eq:prod_strong}), once more the turbulence stress tensor on
the left-hand side was evaluated by means of the enhanced viscosity
approximation.

\begin{figure}
  \begin{center}
    \includegraphics[width=0.95\linewidth]{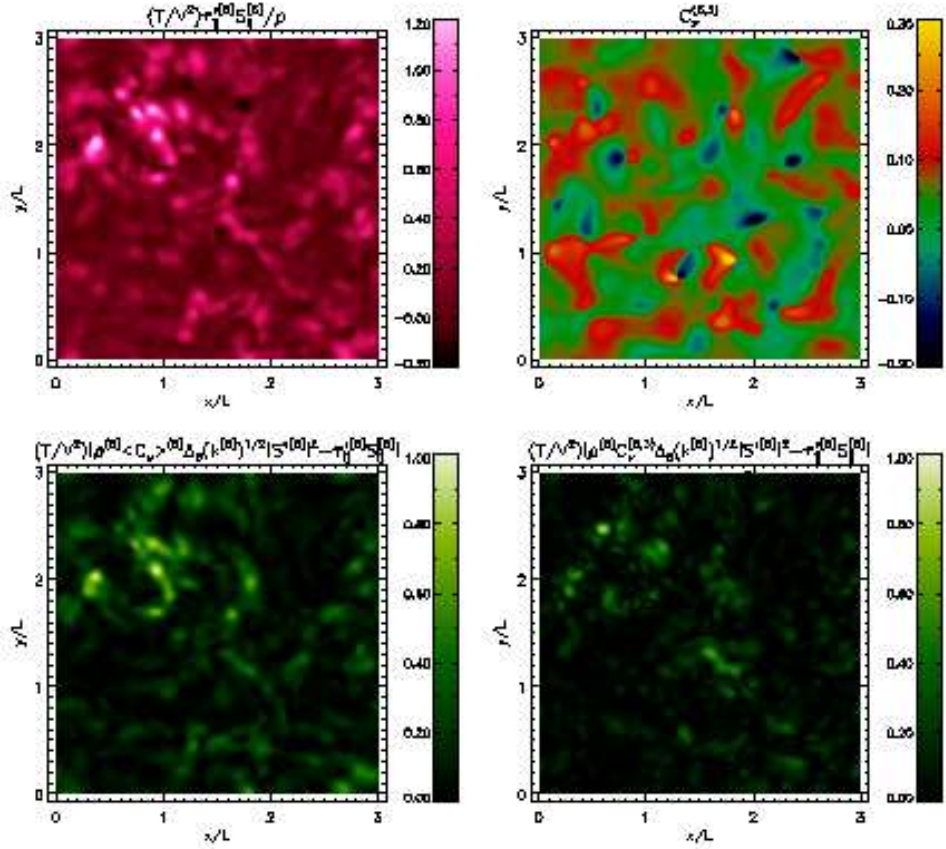}
    \caption{ The rate of turbulence production evaluated from
      filtered data of the simulation with purely solenoidal forcing
      and $V/c_{0}\approx 0.42$ in the plane $z=0$ at time
      $\tilde{t}=2.5$ (a: left top panel) and the corresponding
      deviations of the statistical and the localised
      turbulent-viscosity closure, respectively (c, d: bottom panels).
      Also shown is the localised production parameter inferred from a
      similarity hypothesis (b: right top panel). }
    \label{fg:prod_ersl}
  \end{center}
\end{figure}

The deviations of both the statistical and the similarity closure from
the explicitly evaluated rate of turbulence energy production is
illustrated in figure~\ref{fg:prod_ersl}.  First of all,
figure~\ref{fg:prod_ersl} (a) shows a contour plot of
$\tau^{\ast\,[6]}S_{ik}^{[6]}$, which was computed with a Gaussian
filter of characteristic length $\Delta_{6}=L/6$ from a 2D section of
the flow in the simulation with purely solenoidal forcing and
characteristic Mach number $V/c_{0}\approx 0.42$ at time
$\tilde{t}=2.5$. The corresponding turbulent-viscosity closure with
the production parameter set equal to $\langle
C_{\mathrm{\nu}}^{(6)}\rangle\approx 0.0615$ is plotted in
figure~\ref{fg:prod_ersl} (c). Although the overall agreement is
satisfactory, regions of pronounced production are evidently not
reproduced. On the other hand, the similarity parameter
$C_{\mathrm{\nu}}^{(6,3)}$ was computed
from~(\ref{eq:c_prod_locl}). The corresponding contour plot is shown
in figure~\ref{fg:prod_ersl} (b). Obviously,
$C_{\mathrm{\nu}}^{(6,3)}$ is negative in several regions of the
flow. The resulting negative turbulent viscosity is usually
interpreted as \emph{backscattering} of energy from smaller to larger
scales \cite*[see][\S4.4]{Sagaut}. In fact, energy transfer upwards
through the cascade is well known from turbulence theory and can be
accounted for by localised closures.  The outcome of inserting
$C_{\mathrm{\nu}}^{(6,3)}$ into the turbulent-viscosity closure for
$\tau^{\ast\,[6]}S_{ik}^{[6]}$ is shown in figure~\ref{fg:prod_ersl}
(d). The deviations from explicitly evaluated rate of production are
clearly less than in the case of the closure with the constant
statistical parameter. Although some residual errors persist, the
localised closure performs much better in regions of large turbulence
stresses. Moreover, it gives a fair approximation to backscattering.

\section{The rate of dissipation}

\label{sc:diss}

The rate of dissipation at the level of the $n$-th filter, 
\begin{equation}
  \label{eq:eps_n}
  \epsilon^{(n)} = 
    \nu\langle|\overset{\infty}{S^{\ast}}|^{2}\rangle_{n},
\end{equation}
specifies the rate of conversion of kinetic energy into internal
energy due to microscopic viscous dissipation. The filter operation
corresponds to local averaging of the rate of dissipation over length
scales $l<\Delta_{n}$. Regardless of the filter length $\Delta_{n}$,
the rate of dissipation $\epsilon^{(n)}$ is largely determined by
velocity fluctuations on the smallest dynamical scales near the
microscopic Kolmogorov length $\eta_{\mathrm{K}}$. For this reason,
there is no way of explicitly evaluating $\epsilon^{(n)}$ from
under-resolved numerical data. A simple estimate can be made if the
resolved flow is more or less in stochastic equilibrium. Taking the
global average of the turbulence energy
equation~(\ref{eq:turb_energy}), the diffusion term cancels out, and
the mean time derivative is negligible in the case of steady
turbulence. Hence,
\begin{equation}
  \label{eq:loc_equilbr}
  \left\langle\frac{1}{\rho^{(n)}}\Pi^{(n)}\right\rangle \simeq 
  \langle\epsilon^{(n)}+\lambda^{[n]}\rangle.
\end{equation}
The contribution $\lambda^{(n)}$ stems from pressure dilatation:
\begin{equation}
    \rho^{(n)}\lambda^{[n]} \equiv
    -\langle\overset{\infty}{P}\overset{\infty}{d}\rangle_{n} +
    P^{(n)}d^{[n]}.
\end{equation}
Equation~(\ref{eq:loc_equilbr}) expresses the well known equilibrium
hypothesis for the balance between turbulence production and
dissipation in the quasi-stationary regime for the dynamics
on length scales $l\gtrsim\Delta_{n}$. The validity of this hypothesis
has recently been confirmed for incompressible turbulence by directly
computing the spectral energy transfer function from numerical data
\cite*[see][]{KanIshi03}. 

The most common closure for the rate of dissipation is the dimensional
expression 
\begin{equation}
  \epsilon^{(n)} \circeq 
  C_{\epsilon}^{(n)}\frac{\left\langle\left(k^{[n]}\right)^{3/2}\right\rangle}{\Delta_{n}}.
\end{equation}
In general, the parameter $C_{\epsilon}^{(n)}$ varies in space and
time. However, in the case of statistically stationary homogeneous
turbulence, the closure hypothesis can be simplified further by
assuming a constant value of $C_{\epsilon}^{(n)}$. In order to fulfil
statistical equilibrium, we therefore set
\begin{equation}
  \label{eq:diss_weak}
  C_{\epsilon}^{(n)} = 
  \frac{\Delta_{n}\left\langle\frac{1}{\rho^{(n)}}\Pi^{(n)} - \lambda^{[n]}\right\rangle}
       {\left\langle\left(k^{[n]}\right)^{3/2}\right\rangle}.
\end{equation}
Results calculated from simulation data are shown in
figure~\ref{fg:sgs_stat}.  It appears that the graphs of
$C_{\epsilon}^{(n)}$ flatten near the transition to the range of wave
numbers dominated by dissipation.  In particular, for the simulations
with lower Mach number, $C_{\epsilon}^{(8)}\approx 0.48$.  Even for
the higher Mach numbers, the corresponding values of the dissipation
parameter are not much different. This falls in place very well with
the results discussed by \cite{KanIshi03}.  A further sample of values
for the dissipation parameter is listed in
table~\ref{tb:sgs_compr}. In most cases, $C_{\epsilon}^{(n)}$ is
slightly less than $0.5$ for the filter levels $6\le n\le 12$, which
highlights the robustness of this parameter.

\section{Turbulent diffusion}

\label{sc:diff}

The turbulent flux of kinetic energy is given by
\begin{equation}
  F_{k}^{[n]} = \frac{1}{2}\tau_{iik}^{[n]}+\mu_{k}^{(n)},
\end{equation}
where the contracted third-order moments (TOM) $\tau_{iik}^{[n]}$ and the
pressure-diffusion flux $\mu^{(n)}$ are defined by
\begin{equation}
  \label{eq:tom}
  \tau_{iik}^{(n)} \equiv
  \tau(\overset{\infty}{v_{i}},\overset{\infty}{v_{i}},
       \overset{\infty}{v_{k}}) =
  -\langle\overset{\infty}{\rho}\overset{\infty}{v_{i}}\overset{\infty}{v_{i}}
          \overset{\infty}{v_{k}}\rangle_{n} -
  2\tau_{ik}^{[n]}v_{i}^{[n]} +
  \langle\overset{\infty}{\rho}\overset{\infty}{v_{i}}
         \overset{\infty}{v_{i}}\rangle_{n}v_{k},
\end{equation}
and
\begin{equation}
  \label{eq:press_diff}
  \mu_{k} \equiv \tau(\overset{\infty}{P},\overset{\infty}{v_{k}}) =
  -\langle\overset{\infty}{P}\overset{\infty}{v_{k}}\rangle_{n} +
  P^{(n)} v_{k}^{[n]}.
\end{equation}
In the well known gradient-diffusion closure, the flux is set equal to
the mean product of kinetic diffusivity and the gradient of the
turbulence energy:
\begin{equation}
  \label{eq:grad_diff_vect}
  \vect{F}^{[n]} \circeq 
  \rho^{(n)}C_{\kappa}^{(n)}\Delta_{n}\sqrt{k^{[n]}}\,\vect{\nabla}k^{[n]}.
\end{equation}
This closure immediately reveals a consistency problem. The parameter
$C_{\kappa}^{(n)}$ is over-determined, as there are three flux
components.  For a well defined solution, the gradient of $k^{[n]}$
being aligned with the flux vector $\vect{F}^{[n]}$ is an
indispensable precondition. If this were the case, a least-square
approach could be employed in order to calculate the closure parameter:
\begin{equation}
  \label{eq:diff_weak_lsqr}  
  C_{\kappa}^{(n)} =
  \frac{\langle\vect{F}^{[n]}\cdot\vect{\nabla}k^{[n]}\rangle}
       {\langle\rho^{(n)}\Delta_{n}\sqrt{k^{[n]}}|\vect{\nabla}k^{[n]}|^{2}\rangle}.
\end{equation}

\begin{figure}
  \begin{center}
    \includegraphics[width=0.95\linewidth]{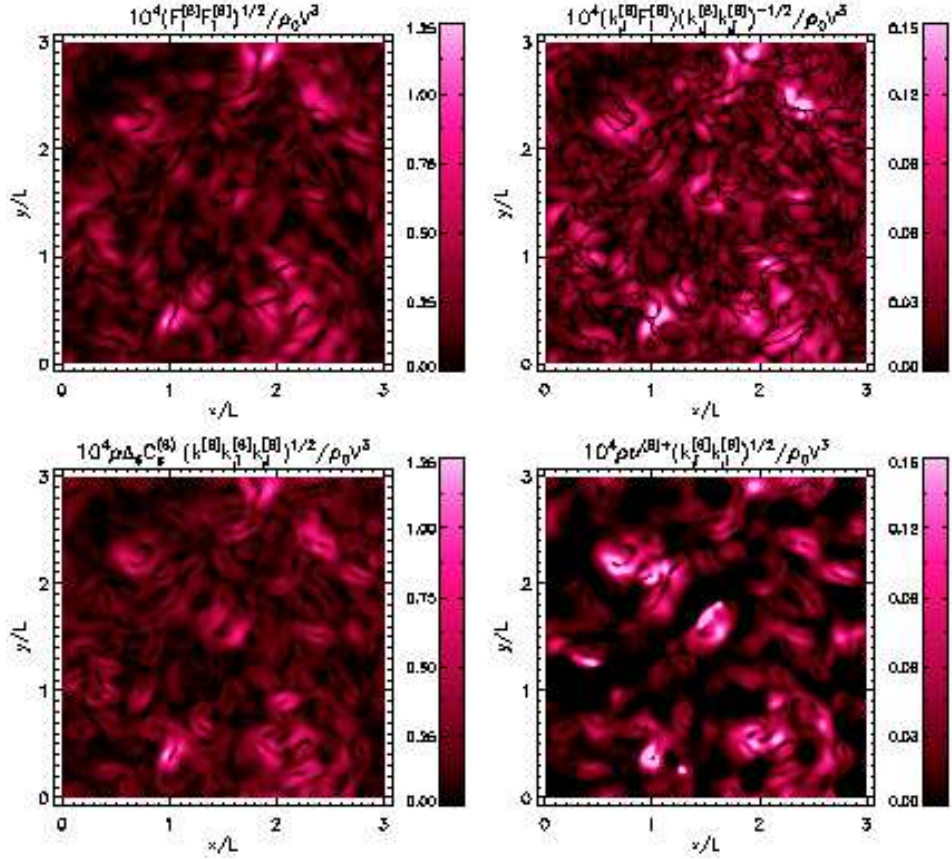}
    \caption{ The turbulent flux magnitude computed with a Gaussian
      filter of characteristic length $\Delta_{6}=L/6$ from data of
      the simulation with $\zeta=0.75$ and $V/c_{0}\approx 0.66$ at
      time $\tilde{t}=5.0$.  Shown are the contours of the actual flux
      magnitude $|\vect{F}^{[6]}|$ (a: left top panel), the projection
      of the modelled flux as defined by
      equation~(\ref{eq:diff_weak_lsqr}) unto the actual flux (b: right
      top panel), and contours of differently modelled fluxes
      (c, d: bottom panels).  }
    \label{fg:diff_flux}
  \end{center}
\end{figure}

In the particular case of the simulation with partially dilatational
forcing ($\zeta=0.66$), the outcome of this ansatz was matched against
the explicitly evaluated flux $F_{k}^{[6]}$ for a 2D section of the
filtered flow at time $\tilde{t}=4.0$. The filtering length is
$\Delta_{\mathrm{6}}=L/6$.  Equation~(\ref{eq:diff_weak_lsqr}) yields
$C_{\kappa}^{(6)}\approx 0.0358$ from the full 3D data set, which is
of the same order of magnitude as the corresponding mean parameter of
production, $C_{\kappa}^{(6)}\approx 0.0574$.  From this, a turbulent
kinetic Prandtl number close to one seems to be supported. However,
the overall agreement between the modelled and the explicitly
evaluated turbulent flux is actually very poor, as is revealed by
a comparison of the figures~\ref{fg:diff_flux} (a)
and~\ref{fg:diff_flux} (b). The flux magnitude is
underestimated by almost a decade, according to the different scales
of either plots.  This discrepancy definitely indicates that the
\emph{a priori} assumption of turbulent transport in the direction of
the turbulence energy gradient is not fulfilled.  In particular, those
points where the turbulence energy gradient is oriented nearly
perpendicular to the turbulent flux vector appear as dark wiggling
ribbons in figure~\ref{fg:diff_flux} (b).

In consequence, we forsook the approach outlined so far and determined the
diffusivity parameter by matching the flux magnitudes rather
than the flux vectors:
\begin{equation}
  \label{eq:diff_weak}
  C_{\kappa}^{(n)}
     = \frac{\langle|\vect{F}^{[n]}|\rangle}
            {\langle\rho^{(n)}\Delta_{n}\!\sqrt{k^{[n]}}|\vect{\nabla}k^{[n]}|\rangle}.
\end{equation}
Implementing this equation, $C_{\kappa}^{(6)}\approx 0.390$ was
obtained from the aforementioned simulation data. In fact, this value is
larger by about an order of a magnitude than the corresponding value
computed with~(\ref{eq:diff_weak_lsqr}). Contours of the
corresponding modelled flux,
$\rho^{(6)}C_{\kappa}^{(6)}\Delta_{6}\sqrt{k^{[6]}}|\vect{\nabla}k^{[6]}|$,
are plotted in 
figure~\ref{fg:diff_flux} (c). The remarkably good correlation to
$|\vect{F}^{[6]}|$ is evident. Even surfaces at which the flux
vanishes, as a result of pressure-diffusion cancelling the TOM
contributions, are well reproduced. Therefore, we conclude that the
gradient-diffusion closure is a fair statistical description of
turbulent transport, which correctly accounts for the magnitude
but not for the local direction of transport.  Numerically calculated
values of $C_{\kappa}^{(6)}$ for different simulations are listed in
table~\ref{tb:sgs_compr}. It appears that there is a trend towards
stronger diffusion for higher Mach numbers.  Figure~\ref{fg:sgs_stat}
shows the variation of $C_{\kappa}^{(n)}$ with the smoothing length
scale $\Delta_{n}$. As one can see, the parameter of diffusion is
almost scale-invariant for $n\in{2,\ldots,10}$, which corresponds to
the nearly inertial subrange.

In addition, we tested whether turbulent diffusivity is correlated
with the turbulent viscosity, since
$\kappa^{(n)}=\sigma_{\mathrm{kin}}^{(n)}\nu^{(n)}$ with
$\sigma_{\mathrm{kin}}\sim 1$ is a commonly used hypothesis.  To this
end, the parameter $C_{\nu}^{(n)}$ was computed locally, according to
the prescription in \S\ref{sc:prod}. Then the diffusivity was set
equal to the viscosity for positive values of $C_{\nu}^{(n)}$.  On the
other hand, $C_{\nu}^{(n)}$ was set identically zero in regions of
negative production, because concentration of turbulence energy due to
negative diffusivity has to be inhibited. The numerical evaluation
demonstrated that the resulting diffusive flux is typically too small
by an order of a magnitude. An exemplary 2D contour section is shown
in figure~\ref{fg:diff_flux} (d).  The assumption of a kinetic Prandtl
number close to unity is thus clearly invalidated.

\section{Conclusion}

We have investigated parameters associated with the common closures
for the turbulence energy budget equation, using data from direct
numerical simulations of compressible isotropic turbulence driven by
stochastic forcing. The statistical self-similarity of isotropic
turbulence, which becomes manifest in the scale-invariance of averaged
closure parameters within the inertial subrange, just begins to emerge
at the resolution of $432$ grid cells per dimension.  Nevertheless, it
was feasible to infer the estimates $\langle C_{\nu}\rangle\approx 0.06$, 
$C_{\epsilon}\approx 0.5$ and $C_{\kappa}\approx 0.4$ for the
parameters of turbulence production, dissipation and diffusion,
respectively. These estimates appear to be robust for Mach numbers
$\lesssim 1$ and predominantly solenoidal forcing. Significant
deviations from these values are likely to emerge from supersonic
turbulence, although we are not able to make definite statements on
the basis of our data. Whereas the results for the parameters of
production and dissipation, respectively, agree well with various
values cited in the literature, we propose a substantially enhanced
parameter of diffusion. The would-be smallness of $C_{\kappa}$, as
given by the standard hypothesis $C_{\kappa}\approx C_{\nu}$, can be
attributed to the mismatch arising from the usually presumed vector
alignment in the gradient-diffusion closure.  We emphasise that there
is no \emph{a priori} reason for any tight relationship between the
turbulence viscosity (specifying the quasi-local transport of energy
from larger to smaller scales) and the turbulence diffusivity
(specifying the non-local redistribution of energy on the smaller
scales).

Furthermore, we compared the modelled turbulence stresses for a
localised eddy-viscosity closure to the explicitly evaluated stresses
and found improved correlation, in contrast to the the closure with a
constant statistical parameter. This result supports the validity of
dynamical procedures for the computation of $C_{\nu}$ in SGS models,
especially, when applying the subgrid scale model proposed by
\cite*{KimMen99}.

For a more stringent analysis of the self-similar regime in numerical
simulations, significantly higher resolution is called for.  Indeed,
extremely high-resolution data have been computed recently
\cite[see][]{KanIshi03}.  These data could be exploited for the
evaluation of structural properties over a wide range of length scales
by means of filtering.

\section{Acknowledgements}

Our implementation of the PPM originated from \cite*{FryMueller89} and
was later adopted in a new code featuring MPI for massively parallel
computation on high-end platforms \cite*[see][]{Rein01} . The
turbulence simulations were run on the Hitachi SR-8000 supercomputer
of the \emph{Leibniz Computing Centre} in Munich, using 512 processors
in parallel. Special thanks goes to M. Reinecke for unfailing
technical advice.  For the post-processing of the data, in particular,
the Gaussian filtering, the \textsl{FFTW} implementation of the the
fast Fourier transform algorithm was utilised
\citep[cf.][]{FriJohn98}. Due to the extraordinary memory
requirements, the computation of the parameters discussed in this
paper was performed on a shared-memory node of the IBM p690
supercomputer of the \emph{Computing Centre of the Max-Planck-Society}
in Garching, Germany. One of the authors (W. Schmidt) was supported by
the priority research program \emph{Analysis and Numerics for
Conservation Laws} of the Deutsche Forschungsgesellschaft.

\bibliography{SimlTurb_JFM}
\bibliographystyle{jfm}

\end{document}